\documentclass[twocolumn,tighten,times]{aastex631}
\usepackage{graphicx}

\definecolor{mygreen}{rgb}{0.475,0.702,0.0259}

\def\gtsim {>\kern-1.2em\lower1.1ex\hbox{$\sim$}~}   
\def\ltsim {<\kern-1.2em\lower1.1ex\hbox{$\sim$}~}   

\received{August 29, 2023}
\revised{December 22, 2023}
\accepted{January 12, 2024}

\begin{document}

\title{Rapid Chemical Enrichment by Intermittent Star Formation in GN-z11}
\author[0000-0002-4343-0487]{Chiaki Kobayashi}
\affiliation{Centre for Astrophysics Research, Department of Physics, Astronomy and Mathematics, University of Hertfordshire, Hatfield, AL10 9AB, UK}
\email{c.kobayashi@herts.ac.uk}

\author[0000-0002-9400-7312]{Andrea Ferrara}
\affil{Scuola Normale Superiore, Piazza dei Cavalieri 7, 50126 Pisa, Italy}

\begin{abstract}
We interpret the peculiar super-solar nitrogen abundance recently reported by the James Webb Space Telescope observations for GN-z11 ($z=10.6$) using our state-of-the-art chemical evolution models. The observed CNO ratios can be successfully reproduced -- independently of the adopted initial mass function, nucleosynthesis yields, and presence of supermassive ($>$1000$M_\odot$) stars -- if the galaxy has undergone an intermittent star formation history with a quiescent phase lasting $\sim$100 Myr, separating two strong starbursts.
Immediately after the second burst, Wolf--Rayet stars (up to $120M_\odot$) become the dominant enrichment source, also temporarily ($<$1 Myr) enhancing particular elements (N, F, Na, and Al) and isotopes ($^{13}$C and $^{18}$O). Alternative explanations involving (i) single burst models, also including very massive stars and/or pair-instability supernovae, or (ii) pre-enrichment scenarios fail to match the data. Feedback-regulated, intermittent star formation might be common in early systems. Elemental abundances can be used to test this hypothesis and to get new insights on nuclear and stellar astrophysics.
\end{abstract}


\section{Introduction} \label{sec:intro}
The James Webb Space Telescope (JWST) is expected to find the first galaxies -- those that host or have hosted metal-free (known as Population III) stars. Surprisingly, though, one of the most distant galaxies detected, GN-z11 at redshift $z=10.6$, showed strong metal lines \citep{bunker23}. Even more puzzlingly, GN-z11 shows an unusually high ($>\!4\times$ solar) N/O ratio.
Super-massive stars have been suggested \citep[e.g.,][]{charbonnel23,senchyna23,nagale23} as the N source in GN-z11. 
Is such an unusual stellar population truly required, or would it be possible to reproduce the observations more simply with a varying star formation history?

At the end of the `dark ages' of the Universe, the cosmic dawn was heralded by the birth of the first stars and galaxies.
The first cosmic star formation is driven by inefficient cooling from hydrogen molecules. Thus, on general grounds, the first stars were expected to be massive, with masses $M_\star\approx100M_\odot$ \citep[e.g.,][]{abel02,bromm04}.
However, the initial stellar mass depends on complex physical processes, such as gas fragmentation, ionization, accretion, and feedback from newborn stars. Once these processes are included in modern numerical simulations, it seems possible to form lower-mass stars \citep{greif11,hirano14, Rossi21}, and even binaries (\citealt{stacy13}; see also \citealt{hartwig23} for observational signatures).

The properties of the first stars, i.e., mass, rotation, multiplicity, and magnetic fields, are important for the reionization and chemical enrichment of intergalactic medium, seeding of super-massive black holes (BHs), and gravitational wave emission.
However, a direct detection of Pop III stars is still lacking, i.e., no zero-metal star or galaxy has yet been found. Instead, the nature of the first stars has been studied using the second generation of stars born out of gas enriched by Pop III stars. 

There is a consensus that second-generation stars can be found among extremely metal-poor (EMP, [Fe/H]\,$<\!-3$) stars in the Milky Way \citep{bee05} and in dwarf spheroidals \citep{Skuladottir21}.
From the analysis of the elemental abundances of EMP stars, it has been deduced that the first enrichment sources were likely to be $M_\star\approx10$--$40M_\odot$ stars, which exploded as `faint' supernovae \citep[e.g.,][]{ume03,ish18}.
Similar results are obtained also for quasar absorption line systems, such as metal-poor damped Lyman-$\alpha$ systems (DLAs), where accurate (barring uncertainties on dust depletion) elemental abundances are measured \citep{kob11dla,saccardi23dla}.

Theoretically, stellar rotation becomes more important at low metallicities, because weaker stellar winds result in a smaller angular momentum loss than at solar metallicity. As a result, if massive stars are fast rotators, rotational mixing brings CNO cycle products into the convective He-burning layers, and the stellar envelope containing light elements such as C, N, and F may be ejected in stellar winds of Wolf--Rayet (WR) stars \citep{mey02,lim18}; 
without them, the N abundance is a dex lower (see Fig.\,9 of \citealt{kob23book}).
WR stars can also explain the detection of highly enhanced HF (hydrogen fluoride) in NGP–190387 at $z=4.42$, a dusty star-forming galaxy discovered by the Atacama Large Millimeter/submillimeter Array \citep[ALMA; ][]{franco21}. 

In addition to NGP–190387, and rather unexpectedly, many high-redshift galaxies contain large ($\simeq10^7M_\odot$) amounts of dust \citep[e.g.,][]{ferrara22, Inami22}. Although supernovae are usually considered the main sources at high-$z$ \citep[$z \sim 7$; ][]{Todini01, dayal22, witstok23}, dust can be produced also by WR stars \citep[e.g.,][]{lau22}.

The earliest JWST observations have revealed an unexpected abundance of super-early ($z>10$), massive ($M_*\,\approx10^9M_\odot$) galaxies at the bright end ($M_{\rm UV}\approx-21$) of the ultraviolet luminosity function. These galaxies tend to have very blue spectral slopes ($\beta<-2.4$). Hence, the dust produced by massive stars associated with the observed stellar population must have been efficiently evacuated (or destroyed) along with most of the gas by powerful galactic outflows driven by the radiation pressure produced by their compact ($\simeq100$ pc), young (20--30 Myr) stellar component \citep{Ferrara23,Ziparo23,Fiore23}. Such feedback temporarily quenches star formation \citep{Looser23, Gelli23} until the gas content of the galaxy is restored by efficient cosmological gas accretion, entailing a duty cycle of $\simeq$50--100 Myr.

In this Letter, we aim to construct a scenario connecting these observations with galactic chemical evolution (GCE) models. As different chemical elements are produced by stars with different masses on different timescales, their abundance ratios can uniquely constrain the star formation and enrichment histories of galaxies.

\begin{table*}
\centering
\begin{tabular}{l|cccccc|ccccr}\hline
& $t_1$ & $t_2$ & $\tau_{\rm i,1}=\tau_{\rm i,2}$ & $\tau_{\rm s,1}$ & $\tau_{\rm s,2}$ & $f_{\rm g,0}$ & $t_{\rm age}^{\rm max}$ & $\left<t_{\rm age}\right>_{\rm M}$ & $\psi$ & $\log$\,O/H\,$+12$ & $\log$\,N/O \\
& [Gyr] & [Gyr] & [Gyr] & [Gyr] & [Gyr] & & [Myr] & [Myr] & [Gyr$^{-1}$] & \\ \hline
\textcolor{mygreen}{Single starburst} & - & - & 0.001 & 0.0002 & - & 0 & 4.2 & 3 & 22 & 7.787 & $-2.198$ \\
\textcolor{red}{Dual starburst (fiducial)}& 0.1 & 0.2 & 0.001 & 0.2 & 0.0002 & 0 & 204 & 26 & 20 & 7.850 & 0.246\\ 
Dual starburst & 0.05 & 0.1 & 0.001 & 0.2 & 0.0002 & 0 & 103 & 9 & 28 & 7.754 & $-0.033$\\ 
Dual starburst & 0.1 & 0.1 & 0.001 & 0.2 & 0.0002 & 0 & 103 & 11 & 38 & 7.754 & 0.047\\ 
Dual starburst & 0.1 & 0.2 & 0.01 & 0.2 & 0.0002 & 0 & 204 & 34 & 41 & 7.769 & 0.208\\ 
Dual starburst & 0.1 & 0.2 & 0.001 & 0.2 & 0.001 & 0 & 205 & 27 & 33 & 7.864 & $-0.081$\\ 
\textcolor{blue}{Pre-enrichment} & - & - & 0.001 & 0.0002& - & 0.3 & 2.8 & 2 & 58 & 7.768 & $-0.374$\\ \hline
GN-z11 & & & & & & & $19_{-5}^{+10}$ & $10_{-2}^{+3}$ & $19_{-12}^{+23}$ & $7.82$ & $>\!-0.25$\\ \hline
\end{tabular}
\caption{Input parameters describing the adopted star formation histories shown in Figs.\,\ref{fig:main} and \ref{fig:xfe}. 
Here, $t_1$ and $t_2$ are the beginning (end) of the first burst in dual burst models.
The formation epoch $t_{\rm age}^{\rm max}$ is chosen to match the observed O/H.
The last four columns are model output values at the observed epoch ($z=10.6$): mass-weighted age of stars $\left<t_{\rm age}\right>_{\rm M}$, star formation rate $\psi$, mass-weighted O abundance and N/O ratio.
The observed values are taken from \citet{bunker23}.
}
\end{table*}

\section{Models}
\subsection{Galactic chemical evolution models}
\label{sec:gce}
%
%
\begin{figure}\center
\includegraphics[width=8.5cm]{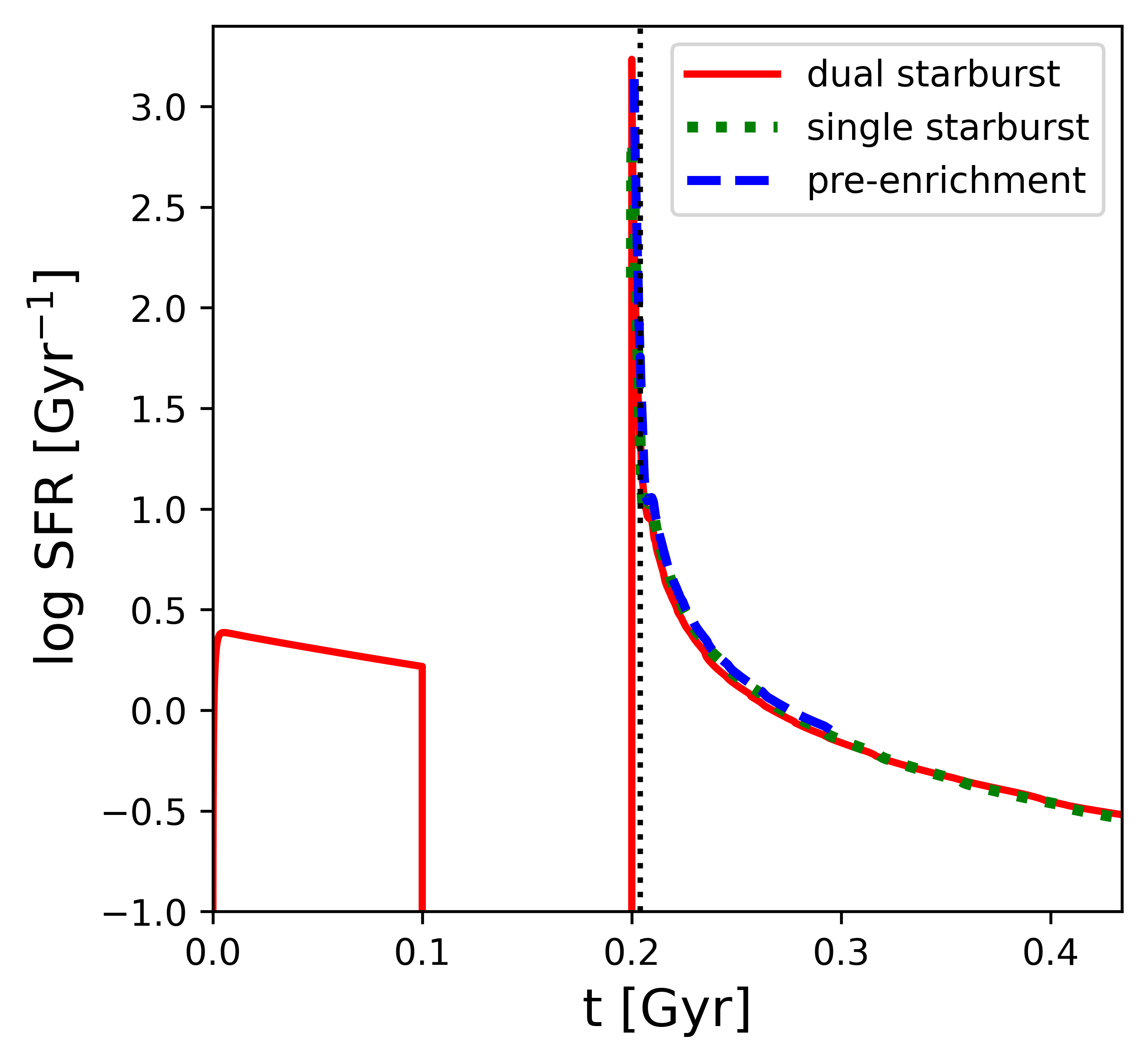}
\caption{\label{fig:sfr}
Star formation history adopted by three different GCE models for GN-z11: single starburst (green short-dashed line); dual starburst (red solid); and single burst with pre-enrichment (blue long-dashed).
The vertical dotted line denotes the observed epoch of GN-z11.
}
\end{figure}
%
%
\begin{figure*}\center
\includegraphics[height=6.5cm]{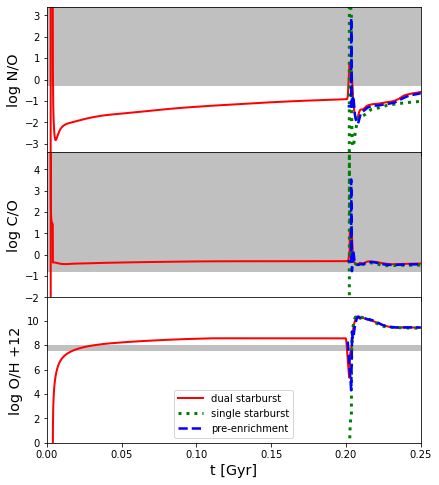}
\includegraphics[height=6.5cm]{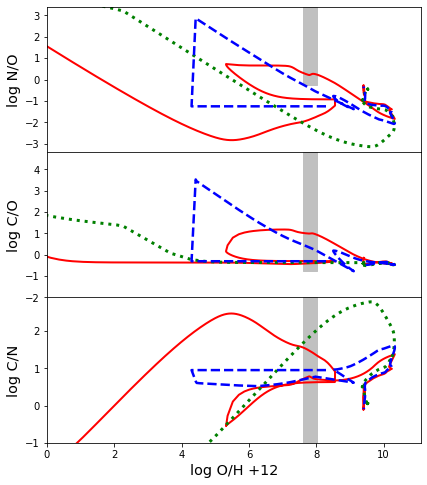}
\includegraphics[height=6.5cm]{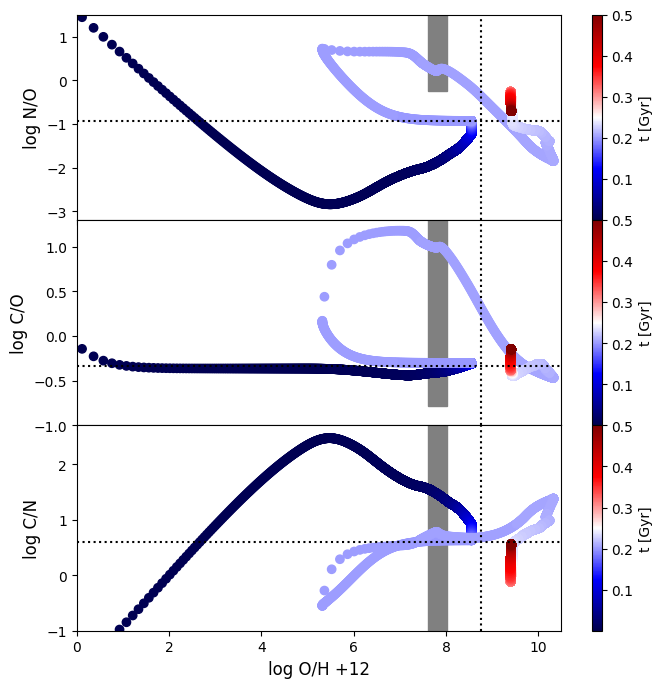}
\caption{\label{fig:main}
Evolution of CNO abundance ratios (by number), compared to the observational data for GN-z11 \citep[][gray areas/bars]{cameron23}.
{\it Left panels}: as a function of time, for the same models in Fig.\,\ref{fig:sfr} including WR stars and standard IMF.
{\it Middle panels}: similar to the left panels but as a function of gas oxygen abundance.
{\it Right panels}: same as the middle panels but only for our fiducial model, i.e., dual starburst with standard IMF, color-coded with the time (in Gyr) elapsed since the first star formation episode.
The dotted lines indicate the solar ratios.
}
\end{figure*}

We use the GCE code from \citet{kob00}, implementing the so-called one-zone model, which integrates the following equation:
\begin{eqnarray}\label{eq:gce}
\frac{d(Z_if_{\rm g})}{dt}&=&E_{\rm SW}+E_{\rm SNcc}+E_{\rm SNIa}\nonumber \\
&-&Z_i\psi+Z_{i,{\rm inflow}}R_{\rm inflow}-Z_iR_{\rm outflow},
\end{eqnarray}
where the mass fraction $Z_i$ of each element $i$ in gas-phase ($f_{\rm g}$ denotes the gas fraction\footnote{Ratio of gas mass to the total mass provided by gas, stars in the galaxy, and the gas ``reservoir'', i.e. $\int_0^\infty R_{\rm inflow} \,dt+f_{\rm g,0}=1$, where $f_{\rm g,0}$ is the initial gas fraction.}) increases via element production ($E_{\rm SW},E_{\rm SNcc}$, and $E_{\rm SNIa}$); the other terms are star formation ($\psi=f_{\rm g}/\tau_{\rm s}$), gas inflow $R_{\rm inflow}=\exp(-t/\tau_{\rm i})/\tau_{\rm i}$, and outflow ($R_{\rm outflow}$) rates, respectively. The model assumes instantaneous mixing of the elements but not instantaneous recycling. 
A complete description can be found in \citet{kob00} and \citet[][hereafter KT23]{kob23book}.
The adopted parameters are summarized in Table 1.

The code includes the latest nucleosynthesis yields of asymptotic giant branch (AGB) stars, super-AGB stars, and core-collapse supernovae (including hypernovae and failed supernovae) from \citet[][hereafter K20]{kob20sr}, as well as WR stars described in \S\,\ref{sec:yields}.
Type Ia supernovae are also included ($E_{\rm SNIa}$) using the model in \citet{kob09} and yields from \citet{kob20ia}. However, their contribution becomes important only at [Fe/H] $\gtsim-1$.

The compositions of the infalling gas ($Z_{i,{\rm inflow}}$) is set to be primordial (see \S\,2.2.1 of K20). All models presented in this Letter have no stars at the start. Our fiducial model assumes initial gas fraction $f_{{\rm g},0}=0$, but we also present a pre-enriched model with $f_{{\rm g},0}=0.3$ and initial composition ($Z_{i,0}$) deduced from another GCE model.

The initial mass function (IMF), taken from \citet{kro08}, is a broken power law in three mass ranges. Later in \S 4, we vary the massive-end slope $x$, and lower and upper mass limits [$m_\ell,m_{\rm u}$] for the Population I/II and III stars, separately. These two IMFs are switched at the threshold (absolute) metallicity $Z_{\rm th}=0.0001$.

It is important to note that our adopted nucleosynthesis yields reproduce the observed elemental abundances in the Milky Way with the standard IMF: $x=1.3,m_\ell=0.01M_\odot$, and $m_{\rm u}=120M_\odot$. This is fundamentally different from the recent modeling works by \citet{isobe23,Marques-Chaves23,bekki23}.

\section{Results}

\subsection{Single burst models}
We first assume a single starburst starting 4 Myr before the observed epoch; this is because the estimated age of GN-z11 stellar populations is $\sim$10 Myr \citep{bunker23}. The corresponding star formation rate per unit mass is shown as a green short-dashed line in Figure \ref{fig:sfr}.

The left panels of Figure \ref{fig:main} show the elemental abundance ratios of the interstellar medium (ISM) predicted by the GCE models as a function of time, while the middle panels
$\log$\,O/H\,$+12$, a commonly used proxy for metallicity\footnote{The solar oxygen abundance adopted in K20 is 8.76.}.
At low metallicities, WR stars produce very high (C,N)/O ratios, which quickly decrease due to the large O production from supernovae.
The N/O ratio increases again at high metallicities because of `primary' production of N from AGB stars, as well as `secondary' production of N (from initially existing CNO) in massive stars.
For the same reason, the initial C/N ratio increase is followed by a rapid drop at high metallicities.

In general, a shorter $\tau_{\rm s}$ shifts the CNO tracks to the right in the middle panels. However, the adopted $\tau_{\rm i}$ and $\tau_{\rm s}$ values (Table 1) are very short, compared with those for present-day massive galaxies. Even shorter $\tau_{\rm s}$ values do not affect the solution any further.
The CNO tracks are insensitive to $\tau_{\rm i}$.
In conclusion, although N/O can be high at very low or very high metallicities, with a single starburst it is not possible to reproduce the observed ratio of GN-z11 at the observed oxygen abundance \citep[7.82; ][]{cameron23}.

\subsection{Dual burst models}
We have seen that WR stars can potentially produce high (C,N)/O ratios, but only at metallicities much lower than observed.
However, there is a way to overcome the problem, and this consists of assuming a dual starburst model.
We now assume that star formation starts\footnote{Results are insensitive to the formation epoch, and thus pre-enrichment from Pop III stars formed at $z\gtsim20$ is included in this subsection.} at $z\sim16.7$ and continues for 100 Myr (Fig.\,\ref{fig:sfr}); the infall and star formation timescales are set to $\tau_{\rm i}=0.001$ and $\tau_{\rm s,1}=0.2$ Gyr, respectively\footnote{The allowed range is $\tau_{\rm s,1}=0.1$--1 Gyr, so that the ISM is sufficiently enriched by the initial star formation. Results are insensitive to $\tau_{\rm i}$.}.

The results of this fiducial model (red curve) are shown in the left and middle panels of Figure \ref{fig:main}.
After the first burst the metallicity already reaches $\log$\,O/H\,$+12=8.5$, and the evolutionary tracks up to this point are similar to those for the single burst case.

Then, star formation is assumed to be completely quenched for 100 Myr\footnote{A marginally consistent, lower-quality fit can nevertheless be obtained also without a quiescent interval between the two bursts.}, possibly due to feedback associated with the onset of an outflow.
At $t=0.2$ Gyr, the secondary infall ($\tau_{\rm i}=0.001$ Gyr) of primordial gas occurs, which initially causes dilution reducing the metallicity.
As a more extreme, second starburst ($\tau_{\rm s,2}=0.2$ Myr)\footnote{The condition to reach the observed N/O is $\tau_{\rm s,2}\le1$ Myr.} is triggered, WR stars quickly enhance (C,N)/O ratios until supernovae produce a large amount of O. The O abundance peaks at $t=0.207$ Gyr, which corresponds to the lifetime of $\sim30M_\odot$ stars, and gradually decreases until $t=0.23$ Gyr. This, along with secondary N production from metal-rich SNe and primary N production from AGB stars (lifetime $\sim$40--150 Myr), leads to a final N/O increase.

This N/O evolution predicted by the dual burst fiducial model crosses the observed range only once, at $t=0.204$ Gyr, i.e., 4 Myr after the onset of the second burst (the right panels of Figure \ref{fig:main}).
The time spent in the data box is very short ($\sim$0.6 Myr). Hence, we conclude that N-enriched objects similar to GN-z11 might be rare.
On the other hand, lower ratios than in GN-z11 (e.g., $\log$\,N/O $=-0.4$ at $z\sim6$ in \citealt{isobe23})
can be reproduced in our scenario with longer $\tau_{\rm s,2}$. In this case, the evolution is slower and these objects become more common.

At the observed epoch, the star formation rate in the fiducial model is 20 Gyr$^{-1}$. This is in excellent agreement with the observed value $\sim$19$M_\odot$ yr$^{-1}$, provided the total stellar mass is $M_\star\sim10^9M_\odot$.
The amount of stars formed during the first burst is only 16\% of the final stellar mass. The mass-weighted age is 26 Myr, which is comparable to the estimated value of $\sim$10 Myr. With shorter first bursts and/or shorter quiescence intervals, the mass-weighted age could be made as short as $\sim$10 Myr.

\subsection{Pre-enrichment?}
An alternative way to change the gas metallicity before the first starburst is by pre-enrichment from external galaxies.
This model is shown in Figure \ref{fig:main}, the left and middle panels (blue long-dashed curve), where it is also compared with the single and dual burst cases.

The pre-enriched model assumes the same $\tau_{\rm i}$ and $\tau_{\rm s}$ as in the single burst one (Table 1), but with a gas chemical composition, $Z_{i,0}$, taken from another (independent) GCE model with $\tau_{\rm i,0}=1$ and $\tau_{\rm s,0}=0.3$ Gyr. We set the initial gas fraction $f_{\rm g,0}=0.3$.
These $f_{\rm g,0}$ and $Z_{i,0}$ values are very similar to those obtained for the dual starburst model during the interval after the first burst, i.e., at $t=0.1$--0.2 Gyr.
However, in this case, the initial stellar fraction is zero, which is the key difference from the dual burst model (red curve).

The pre-enrichment model starts at $\log$\,O/H\,$+12=8.3$. Due to the dilution with the infalling pristine gas, the metallicity (horizontally) decreases keeping the same CNO ratios. The (C,N)/O ratios quickly (vertically) increase due to enrichment from WR stars, and they return to their initial composition values.
Hence, the triangle track does not cross the observed range. The above results are insensitive to the parameters of $\tau_{\rm i}$ and $\tau_{\rm s}$. By changing the initial gas fraction $f_{\rm g,0}$ and/or the initial composition $Z_{i,0}$, we cannot find any tracks that match the observed abundance range. We conclude that this simple pre-enrichment prescription cannot solve the problem at hand. 

\section{IMF dependence}
%
%
\begin{figure*}\center
\includegraphics[height=9.9cm]{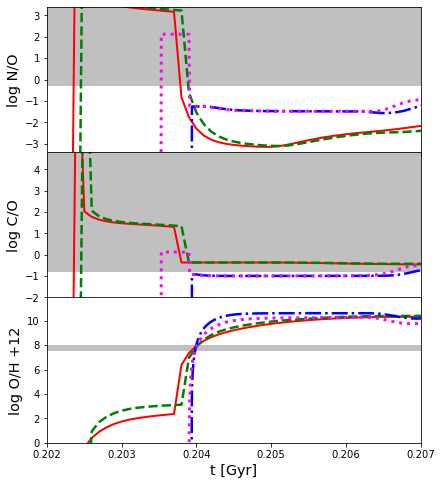}
\includegraphics[height=9.9cm]{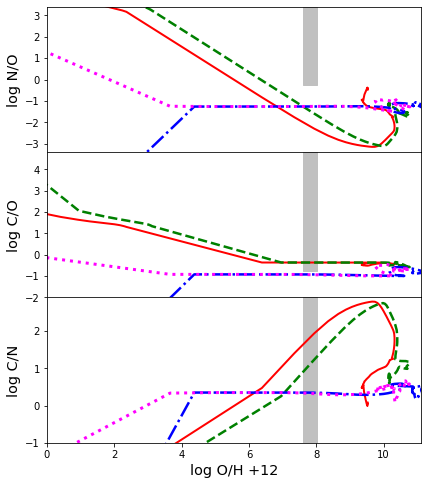}
\caption{\label{fig:dd}
Evolution of CNO abundance ratios (by number) vs. time (left) and gas oxygen abundance (right) for
single starburst models with different IMFs.
The standard model includes WR stars assuming Kroupa IMF with massive-end slope of $x=1.3$ (red solid lines).
The additional curves assume instead a top-heavy IMF for Pop III stars with a slope $x=0$ in the mass range $30$--$120M_\odot$ (green short-dashed), $100$--$280M_\odot$ to add PISNe (blue long-dashed), and $100$--$1000M_\odot$ to add VMSs (magenta dotted). 
Gray areas/bars are the observational data for GN-z11 \citep{cameron23}.
}
\end{figure*}

\subsection{Fate of massive stars}
\label{sec:yields}

While the ultimate fate of stars depends on their initial mass\footnote{With the term \textit{initial mass} we refer to single stars, as stars can lose their envelope also due to binary interactions.}, the final mass and nucleosynthesis yields are determined by stellar mass-loss \citep[e.g.,][]{vink11} as well as convection treatment and nuclear reaction rates.
We include nucleosynthesis yields covering a full mass range up to $1000M_\odot$ as follows.

10--50$M_\odot$: These stars become core-collapse supernovae. Although \citet{lim18} provided yields including explosive nucleosynthesis, their yields do not match observations in the Milky Way, probably because of the lack of hypernovae and mixing fallback. Therefore, we combine their stellar wind yields (see below) with the explosive nucleosynthesis yields of CO cores from K20, which result in an excellent agreement with the observations of almost all elements\footnote{At $Z=0$, some stars in the range 13--40$M_\odot$ may explode as `faint' supernovae/hypernovae.
These stars produce only a tiny amount of iron because of the relatively large remnant BH ($\sim$5$M_\odot$). Their contribution is negligible and is not included here.}.
Roughly half of stars with 20--50$M_\odot$ are assumed to explode as hypernovae leaving a BH, probably due to rotation and/or binary interaction. The exact fraction of hypernovae can be determined from chemodynamical simulations of a Milky Way-type galaxy; \citet{kob11mw} find $\epsilon_{\rm HN}=(0.5,0.5,0.4,0.01,0.01)$ for $Z=(0,0.001,0.004,0.02,0.05)$ . All stars with 13--20$M_\odot$, and the rest of stars with 20--30$M_\odot$, explode as normal core-collapse supernovae with $10^{51}$ erg of explosion energy, leaving a neutron star. The rest of stars with 30--50$M_\odot$ are assumed to be `failed' supernovae, which form a $\sim$10$M_\odot$ BH. This is based on the unsuccessful explosion simulations of such massive stars \citep{jan12,bur21} and the lack of massive progenitors expected at supernova locations in HST data \citep{sma09}, but the threshold was determined from GCE models (Fig.\,4 of K20).

60--140$M_\odot$: Stars above $\sim$90$M_\odot$ become \textit{pulsating pair-instability supernovae}. Their evolution and nucleosynthesis are uncertain, but they will leave a $\sim$100$M_\odot$ BH. Although this phase is not calculated, we use the stellar wind yields from \citet{lim18} for 13--120$M_\odot$ with four different metallicities and three rotational velocities. With rotation, lower-mass stars become WR stars.
We apply rotating models to a metallicity-dependent fraction $\epsilon_{\rm HN}(Z)$ (given above) of the stars; among these, 3\% (97\%) rotate at 300 (150) km s$^{-1}$. The remaining $1-\epsilon_{\rm HN}(Z)$ fraction of stars are assumed to be non-rotating.

160--280$M_\odot$: If the masses of the first stars are $\sim$160--280$M_\odot$, they explode as \textit{pair-instability supernovae} \citep[PISNe; ][]{barkat67,heg02,nom13,tak18} leaving no remnant. PISNe have a very distinct nucleosynthetic pattern. Although considerable effort has been made to detect such a characteristic pattern, no observational signature for the existence of PISNe has yet been convincingly found, neither in EMP stars \citep{cay04,Aguado23} nor in DLAs \citep{kob11dla,saccardi23dla}.
We apply a mix of non-magnetic, rotating and non-rotating models from \citet{tak18} at $Z=0$ only, using $\epsilon_{\rm HN}(Z)$ described above.

300--1000$M_\odot$: \textit{Very-massive stars} (VMSs; $>$100$M_\odot$) have been solidly identified in the Tarantula Nebula of the Large Magellanic Cloud \citep{schneider18}.
Various stellar evolution models \citep[e.g.,][]{szecsi22}, and some nucleosynthesis yields \citep{yusof13,martinet22,volpato23} exist, although the results significantly depend on the input physics.
We take the nucleosynthesis yields of the `max' mass-loss models from \citet{volpato23} for 300--1000$M_\odot$ and 100--1000$M_\odot$ at metallicity $Z=0$ and $0.0002$, respectively.
We also take the stellar wind yields of the V11 models from \citet{higgins23} for 100--500$M_\odot$ at $Z=0.014$, 
assuming very strong winds from \citet{vink11}. 
N production is seen as a result of the CNO cycle during core H-burning.
These yields do not include explosive nucleosynthesis (i.e., stellar winds only).

We do not include \textit{super-massive stars} (SMSs; $>$1000$M_\odot$) in our GCE models, as their properties and yields are uncertain.
SMSs are originally defined as those that collapse on the general relativistic instability before igniting H-burning \citep{ful86}.
Incomplete core H-burning of metal-enhanced stars with $>10^4M_\odot$ has been proposed as an explanation for the abundance anomaly \citep{denissenkov14}, and the so-called O--Na anti-correlation \citep{kraft97} often seen in globular clusters of the Milky Way.
These stars have been invoked also to account for the high N/O ratio measured in GN-z11 \citep{nagale23}, although lower-mass stars survive until He-burning and enhance C and O.
However, the structure and evolution depend not only on their initial masses and metallicities, but also their formation path \citep{woods19}.
In a very narrow mass range, zero-metallicity stars may explode as general relativistic supernovae \citep{chen14}, which will have a significant effect and would be inconsistent with the observation of GN-z11.

\subsection{CNO evolution}

The IMF dependence on the CNO tracks are shown in Figure \ref{fig:dd} as a function of time (left) and ISM oxygen abundance (right).
In order to maximize the impact of VMSs, a top-heavy IMF is assumed for Pop III stars with a flat mass distribution ($x=0$).
At low metallicities, compared to the standard IMF (red solid lines), a top-heavy\footnote{We adopt $m_\ell=30M_\odot$ for Pop III to minimize O production from core-collapse supernovae.} Pop III IMF up to $m_{\rm u}=120M_\odot$ (green short-dashed) can produce up to $\sim1$ dex systematic increase of N/O. 
If the Pop III IMF is extended up to $m_{\rm u}=280M_\odot$ (blue long-dashed), PISNe produce a large amount of Fe and also O, and the initial (C,N)/O ratios become very low.
Finally, if the Pop III IMF is extended up to $m_{\rm u}=1000M_\odot$ (magenta dotted), the initial (C,N)/O ratios can become high again due to the VMS contribution (\S \ref{sec:yields}).

In the last two models with $m_{\rm u}>120M_\odot$, while PISNe are assumed to occur only from Pop III stars, VMSs produce some metals via winds also at higher metallicities.
This causes a $\sim2$ dex increase of N at intermediate metallicities ($\log$\,O/H\,$+12\sim8$) which is however insufficient to match the observed range of GN-z11.
Larger $m_{\rm u}$ values do not increase the N/O ratio further because of the Pop I/II IMF slope of $x=1.3$.
The VMS contribution is washed out once AGB stars start producing N at high metallicities. Also note that VMSs decrease C (as it is transformed into N), and these two models do not reproduce the observed C/O ratio either.

\section{Conclusions}

We have presented, for the first time, a chemical evolution model that naturally explains the ``anomalous'' elemental abundance ratios of GN-z11.
Without changing the IMF or the nucleosynthesis yields, the observed data can be reproduced if this galaxy has experienced a star formation history featuring a quiescent phase, lasting $\sim$100 Myr, separating two strong starbursts.
Importantly, the observed ratios cannot be explained by single burst models with VMSs and/or PISNe. Essentially, this is because VMSs do not increase N/O sufficiently, while PISNe rather decrease N/O ratios.

In our successful models, prior to the observed epoch ($z=10.6$), the galaxy has been chemically enriched.
This pre-enrichment is likely caused by internal, rather than external, sources.
For a brief period after the second burst, 
WR stars (up to $120M_\odot$) become the dominant enrichment source, which explains the high (C,N)/O ratios at the observed metallicity of GN-z11. WR stars also enhance particular elements (N, F, Na, and Al; see Appendix) and isotopes ($^{13}$C and $^{18}$O); the prediction of high fluorine abundance can be tested with ALMA \citep[e.g.,][]{franco21}.

Our results strongly suggest that super-early galaxies undergo a feedback-regulated, stochastic \citep{Pallottini23}, or even intermittent \citep{Cole23}, star formation history. In spite of this, detecting the chemical anomalies (e.g. high N/O) produced by multiple bursts might not be easy as the chemical evolution proceeds quite rapidly, thus erasing this signature on short time scales ($\ltsim1$ Myr).
This scenario is also consistent with the presence of strong radiation-driven outflows \citep{Carniani23}, which are necessary to clear the dust produced by the observed stars in super-early galaxies like GN-z11 \citep{Ferrara23, Ziparo23, Fiore23}.
At lower redshifts ($z\sim7$), dust instead may accumulate inside galaxies, which might correspond to dusty galaxies observed by ALMA \citep[e.g.,][]{dayal22}.

Our findings highlight the potential of chemical evolution models and data on elemental abundances (and isotopic ratios), not only to investigate the physics and evolution of the most distant galaxies but also to provide new constraints on nuclear and stellar astrophysics.

\begin{acknowledgments}
We acknowledge the stimulating atmosphere of the IFPU Focus Week (May 2023), where this work has been initially worked out. We thank M. Limongi, K. Takahashi, and G. Volpato for providing the nucleosynthesis data. CK acknowledges funding from the UK Science and Technology Facility Council through grant ST/R000905/1, ST/V000632/1. The work was also funded by a Leverhulme Trust Research Project Grant on ``Birth of Elements''. Support from ERC Advanced Grant INTERSTELLAR H2020/740120 (AF) is kindly acknowledged.
\end{acknowledgments}

\section*{Data Availability}

The data underlying this article will be shared on reasonable request to the corresponding author.

\appendix
\vspace*{-3mm}

\begin{figure}
\center
\includegraphics[height=13.cm]{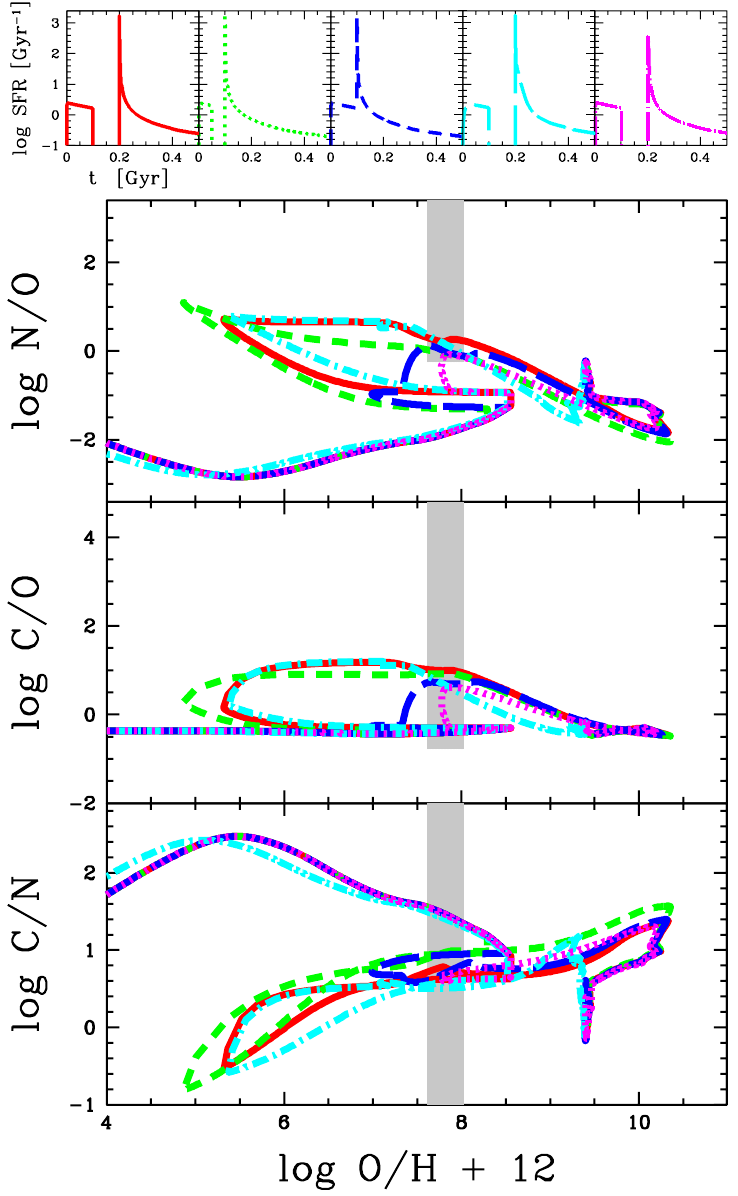}
\includegraphics[height=13.cm]{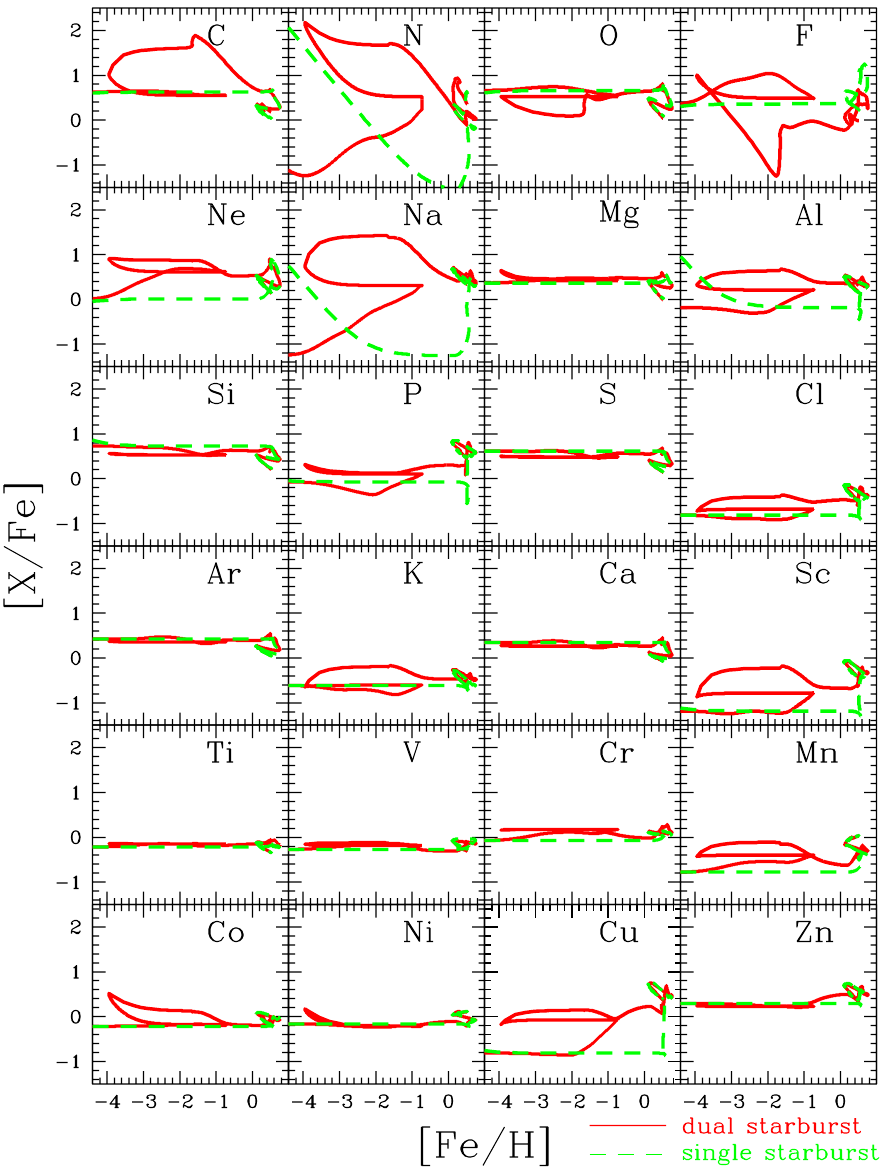}
\caption{\label{fig:xfe}
{\it Left}: same as the middle panel of Fig.\,\ref{fig:main} but for a variety of dual burst models assuming a shorter interval (green short-dashed line), no interval (blue long-dashed), longer $\tau_{\rm i}$ (cyan dot-dashed), and shorter $\tau_{\rm s,2}$ (magenta dotted), compared to the fiducial model (red solid). 
The corresponding star formation histories are shown on the top panels.
{\it Right}: the [X/Fe]--[Fe/H] relations for the dual (red solid line) and single (green dashed) burst models in Fig.\,\ref{fig:main}.
}
\end{figure}

Our conclusions are based on the following parameter study.
The left panels of Figure \ref{fig:xfe} show a variety of dual burst models that can reproduce the observations, compared to the fiducial model (red solid line);
the differences are small depending on the details of the dual burst models.
(i) If the duration of the first burst and/or interval between the two bursts are short ($t_1=0.05,t_2=0.1$ Gyr; green short-dashed), similar CNO tracks are obtained, with a slightly lower peak N/O ratio.
(ii) If no interval ($t_1=t_2=0.1$ Gyr; blue long-dashed) is assumed, the peak N/O ratio is only marginally consistent with data, since the metallicity decrease due to dilution after the secondary gas inflow is weaker.
(iii) Similar CNO tracks are also obtained with longer inflow timescales ($\tau_{\rm i}=0.01$ Gyr), which also result in a higher star formation rate at the observed epoch.
(iv) Finally, with a weaker second burst ($\tau_{\rm s,2}=0.001$ Gyr), the N/O ratio becomes lower, and the condition to reach the observed N/O is $\tau_{\rm s,2}\le1$ Myr.

Our GCE models can predict elemental abundances and isotopic ratios, self-consistently.
The right panels of Figure \ref{fig:xfe} shows the [X/Fe]--[Fe/H] relations of all stable elements from C to Zn, for our fiducial, dual burst model (red solid line) and a single burst model (green dashed). For GN-z11, it would be difficult to measure the abundances of elements other than CNO. Nevertheless, these predictions can be compared with peculiar stars in the Milky Way to study its formation process.

\newpage
\bibliography{ms}{}
\bibliographystyle{aasjournal}

\end{document}